\documentclass[prl,twocolumn,preprintnumbers,tightenlines,superscriptaddress]{revtex4}
\usepackage{graphicx}
\usepackage{amssymb}
\usepackage{amsmath,mathrsfs,verbatim}
\usepackage{graphics}
\usepackage{times}
\usepackage{latexsym}
\newcounter{sec}
\allowdisplaybreaks
\begin{document}

\title{Matter Power Spectrum of Light Freeze-in Dark Matter: With or without Self-Interaction}

\author{Ran Huo}
\affiliation{Department of Physics and Astronomy, University of California, Riverside, California 92521, USA}

\date{\today}

\begin{abstract}
We study the free-streaming effect in a light freeze-in dark matter model. Naturally in the dark sector one can find dark matter related coupling, and such coupling may induce dark matter self-scattering. In case that such scattering is subdominant, the dark matter partition function is not thermal but determined by the freeze-in process, yet its high momentum side is generally also Boltzmann suppressed. We show that the matter power spectrum is very similar to a warm dark matter one in shape. When matched to the current WDM bound, a $24$~keV freeze-in dark matter is ruled out at $2\sigma$ confidence level. In case that the dark matter self-scattering is strong and decouples at a very late time, by a new numerical calculation we show that the early stage Brownian motion indeed protects the power spectrum against free-streaming suppression. However, such an effect cannot be characterized by a free-streaming length alone; we find that the self-scattering decoupling time is another necessary parameter. The currently interested dark matter self-interaction cross section $\sim\text{cm}^2/\text{g}$ is just marginal for such protection to be effective.
\end{abstract}

\maketitle

\stepcounter{sec}
{\bf \Roman{sec}. Introduction.\;}
In the past, Weakly Interacting Massive Particle has played a central role in dark matter (DM) model building. For a particle with weak scale mass and more importantly weak scale interaction with the standard model (SM) particle ($\mathcal{O}(1)$~pb annihilation cross section), the relic abundance $\Omega_\chi h^2=0.1186$~\cite{Aghanim:2018eyx} as the most explicit measurement by far for DM arises naturally by the so-called thermal freeze-out mechanism~\cite{Kolb:1990vq}. However, current direct~\cite{Aprile:2018dbl,Cui:2017nnn}, indirect~\cite{TheFermi-LAT:2017vmf} and collider~\cite{Boveia:2018yeb} searches for such a particle have substantially probed its parameter space, and the null results have reduced the motivation for this mechanism. Alternatively, freeze-in mechanism~\cite{Ellis:1984eq,McDonald:2001vt,Choi:2005vq,McDonald:2008ua,Hall:2009bx} can also parametrically give correct relic abundance, in which DM is not present at the very beginning, but produced through scattering with the visible sector particles (see~\cite{Bernal:2017kxu} and references therein).

Such DM particles naturally inherit the thermal motion of the visible sector particles. For light DM candidate which is becoming more popular either from direct detection perspective~\cite{Knapen:2017xzo,Ibe:2017yqa} or motivated by potential DM indirect detection signal~\cite{Babu:2014pxa,Biswas:2017ait,Heeba:2018wtf} or even more astrophysically~\cite{McDonald:2001vt,deVega:2013jfy,An:2018nvz}, a significant free-streaming (FS) effect is expected. However the freeze-in DM (FIDM. In literature it is often called ``Feebly Interacting Massive Particle'', but here it is neither massive nor necessarily feeble in self-interaction.) is usually taken as nonthermal with its partition function determined in the freeze-in process~\cite{Roland:2016gli,Konig:2016dzg,Biswas:2016iyh,Bae:2017dpt}, so for its matter power spectrum (MPS) strictly one cannot directly use a warm DM (WDM) one which is easily available. On the other hand, in DM model building it is natural to find (even order one) couplings in the dark sector (\emph{e.g.}~see~\cite{McDonald:2001vt}) which induces DM self-scattering~\cite{Spergel:1999mh,Tulin:2017ara}, and such scattering tends to bring the dark sector back into thermal equilibrium. Whether such self-scattering can thermalize the FIDM is up to parameters.

In this \emph{Letter}, we will consider two opposite limits in detail in which the MPS of FIDM is determined differently. In the first case that the self-scattering is ineffective, we will calculate the partition function from the first principle and show that it is also Boltzmann suppressed on the high momentum side, then the MPS difference~\cite{Konig:2016dzg} caused by the difference of the shape of the partition distribution compared to a thermal distribution is in general negligible. In the opposite limit that the self-scattering is very fast, so the FS gives way to Brownian motion early on, we will show that a total FS length is insufficient to characterize the MPS suppression, and a second parameter of the self-scattering decoupling time/scale factor will be necessary.

\stepcounter{sec}
{\bf \Roman{sec}. The Weak Self-Scattering Limit.\;}
While the generic feature revealed by our calculation does not depend on specific model realization, we still use a certain freeze-in model for convenience. Our example model~\cite{Yaguna:2011qn,Babu:2014pxa,Heeba:2018wtf} contains a real scalar gauge singlet extension of the SM Higgs sector
\begin{equation}
\label{eq:Lagrangian}
\mathcal{L}\supset \frac{1}{2}(\partial_\mu\Phi)^2+\frac{1}{2}\mu^2\Phi^2-\frac{1}{4}\lambda_\phi\Phi^4-\frac{1}{2}\lambda_{h\phi}H^\dag H\Phi^2.
\end{equation}
Here $H$ is the SM Higgs doublet, and $\Phi$ can be decomposed into $\phi+u$, the particle and the vacuum expectation value. After both electroweak symmetry breaking and the $u$ turning on, the mixing angle between the SM Higgs and the new $\Phi$ sector $\theta\approx\frac{\lambda_{h\phi}uv}{m_h^2-m_\phi^2}$ is suppressed by the ``feeble'' interaction of $\lambda_{h\phi}$.

This simple model is motivated by the $3.55$~keV X-ray extragalactic line excess, since if $m_\phi=7.1~$keV then decay of $\phi$ through the SM Higgs portal is only kinetically open to photon pairs in the SM. In \cite{Heeba:2018wtf} the freeze-in process is calculated in detail, the post-electroweak phase transition freeze-in production is found to be dominant, and the self-scattering is insufficient to bring the DM sector back into thermal equilibrium through the $2\to3$ process. The simplification here compared to models in~\cite{Konig:2016dzg,Biswas:2016iyh} is that we do not need a new portal, as well as its independent Boltzmann equation.

We start the calculation of partition function with focus on the energy of the outgoing $\phi$. Let $r=\frac{m_\phi}{m_h}$, boosting the 4-momentum of the outgoing $\phi$ from the Higgs rest frame (in which $p_\phi^\mu=(\frac{m_h}{2},~\frac{m_h}{2}(1-4r^2)^{\frac{1}{2}}\hat{r})$ and $\hat{r}=(\sin\theta\cos\varphi,\sin\theta\sin\varphi,\cos\theta)$ is the unit vector in spherical coordinate) into a frame in which the Higgs energy is $E_h$, we get $E_\phi=\frac{1}{2}(E_h+((E_h^2-m_h^2)(1-4r^2))^{\frac{1}{2}}\cos\theta)$. In this frame the spectrum distribution with $E_\phi$ is uniform since $d(\text{probability})/dE_\phi\propto \sin\theta d\theta d\phi/d\cos\theta$ is $\theta$ independent. So for differential partition function in the freeze-in process,
\begin{equation}
\label{eq:differential}
\frac{df_\phi}{dE_\phi}=2\int_{E_{h\text{min}}(E_\phi)}^{E_{h\text{max}}(E_\phi)}\frac{dE_h}{\sqrt{(E_h^2-m_h^2)(1-4r^2)}}\frac{df_h}{dE_h},
\end{equation}
where the factor $2$ counts the two $\phi$s produced in a single decay event, and $E_{h\text{min}}(E_\phi)$ ($E_{h\text{max}}(E_\phi)$) is the minimal (maximal) $E_h$ value to still give a certain $E_\phi$ energy.

In the expanding universe the Boltzmann equation is $\frac{1}{a^3}\frac{d}{dt}\big(a^3f_h\big)=-\frac{m_h}{E_h}\Gamma_{h\to\phi\phi}f_h$,
where $a$ is the scale factor with normalization $a_0=1$, and we have simplified by using $\Gamma_{h\to\phi\phi}=\frac{1}{2m_h}\int\frac{d^3p_{\phi_1}d^3p_{\phi_2}(2\pi)^4\delta^{(4)}(p_h-p_{\phi_1}-p_{\phi_2})}{(2\pi)^62E_{\phi_1}2E_{\phi_2}}\sum|\mathcal{M}|^2$, neglecting the inverse process of freeze-in. Let $x=\frac{m_h}{T}$, then the $x$-based Boltzmann equation reads
\begin{equation}
\label{eq:boltzmann1}
\frac{d}{dx}\frac{f_h}{s}=-\frac{\Gamma_{h\to\phi\phi}}{xH(x)}\frac{m_h}{E_h}\frac{f_h}{s},
\end{equation}
where $s$ is the entropy density. Plugging Eq.~\ref{eq:boltzmann1} into the right hand side of Eq.~\ref{eq:differential} we will get the partition function of $\phi$ determined by the freeze-in process
\begin{equation}
\label{eq:boltzmann2}
\frac{d^2(f_\phi/s)}{dxdp_\phi}=\frac{2(p_\phi/E_\phi)\Gamma_{h\to \phi\phi}}{ xH(x)\sqrt{1-4r^2}}\int_{p_{h\text{min}}}^{p_{h\text{max}}}\frac{dp_h}{E_h^2}
\Big(\frac{m_h}{E_h}+\frac{m_h}{T}\Big)\frac{e^{-\frac{E_h}{T}}}{s},
\end{equation}
where we have used $pdp=EdE$ and Boltzmann distribution for Higgs.

Next let $x_\phi=\frac{p_\phi}{m_\phi}$, $x_h=\frac{p_h}{m_h}$ and $y_\phi=\frac{p_\phi}{T}=\frac{p_\phi}{m_\phi}\frac{m_\phi}{m_h}\frac{m_h}{T}=rx_\phi x$. Apparently $y_\phi\sim1$, $x_h\sim1$ and $x_\phi\sim r^{-1}\gg1$. Note that $y_\phi$ also has the advantage that it is Hubble expansion invariant, it is the right variable for the FS partition function. Then integrating over $x$ we get
\begin{align}
\label{eq:boltzmann3}
\frac{d(f_\phi/s)}{dy_\phi}=&\frac{2\Gamma_{h\to \phi\phi}}{\sqrt{1-4r^2}}\int_0^\infty\frac{y_\phi}{\sqrt{(rx)^2+y_\phi^2}}\frac{dx}{x^2H(x)}\nonumber\\
\times&\int_{x_{hl}}^\infty\frac{dx_h}{1+x_h^2}\Big(\frac{1}{\sqrt{1+x_h^2}}+x\Big)\frac{e^{-x\sqrt{1+x_h^2}}}{s},
\end{align}
where $x_{hl}=\frac{1}{2r}|\sqrt{(1+(\frac{y_\phi}{rx})^2)(1-4r^2)}-\frac{y_\phi}{rx}|$ is the integration lower bound, and the upper bound has been relaxed to infinity with help of Boltzmann suppression of Higgs.

\begin{figure}[t!]
\includegraphics[scale=0.5]{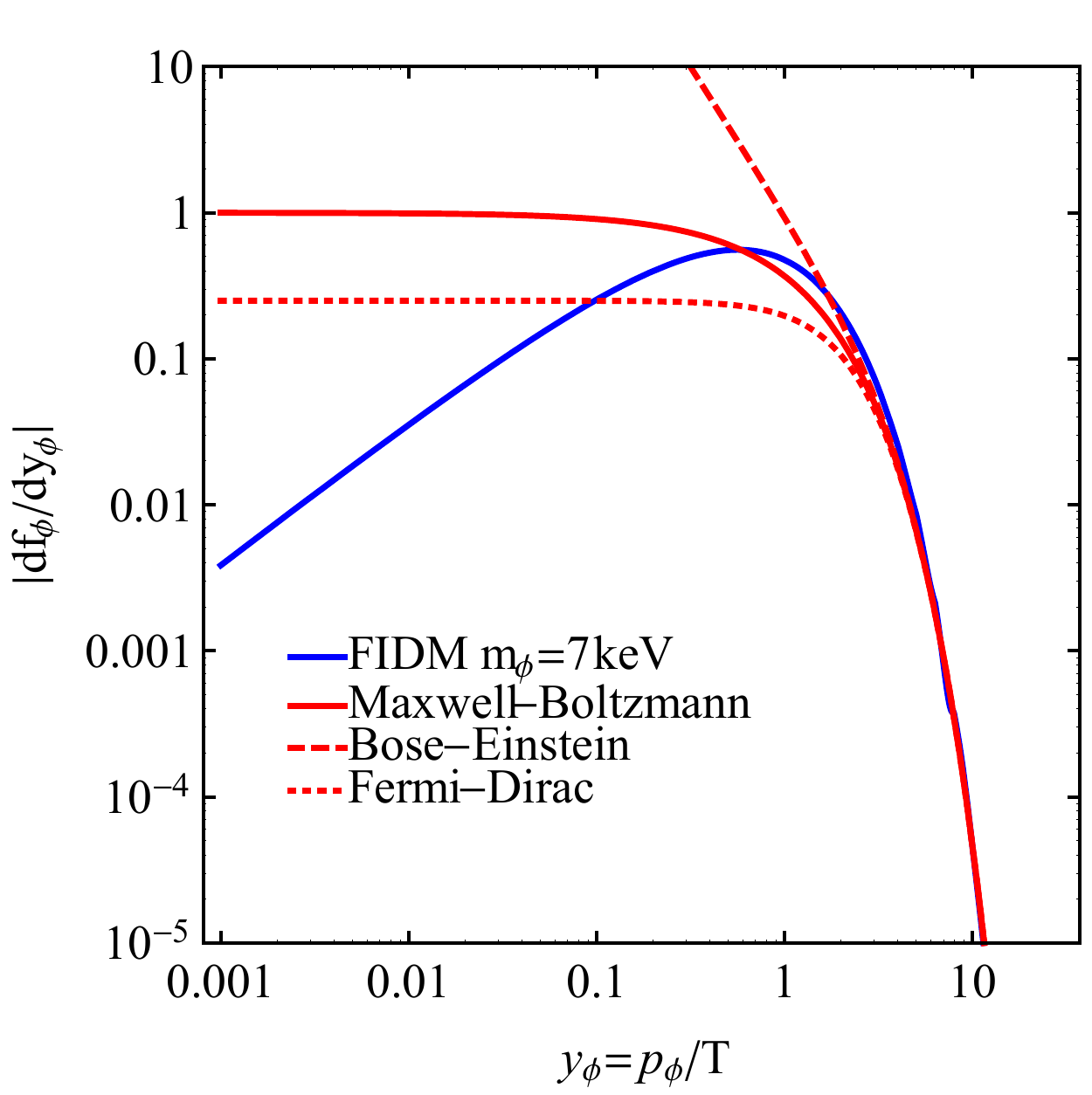}
\caption{Comparison of FIDM differential partition function $|\frac{df_\phi}{dy_\phi}|$ of our model to the thermal equilibrium ones, where $y_\phi=\frac{p_\phi}{T}$. The FIDM partition function is multiplied by a factor so that it is comparable in size to the thermal ones, especially for $y_\phi\gtrsim1$. For thermal distributions, we assume no chemical potential.}
\label{fig:dfdxs}
\end{figure}
Then we can calculate the differential freeze-in partition function numerically by Eq.~\ref{eq:boltzmann3}. Here the reference temperature $T$ can be determined by the entropy conservation $g_{\ast s}(aT)^3=\text{constant}$, with reference values today $g_{\ast s,0}=2+\frac{7}{8}*2*3.046*\frac{4}{11}=3.93$ and $T_0=T_\text{CMB}=2.7255$~K. For arbitrary new physics, $g_{\ast s}$ should be taken as a free parameter. The SM gives $g_{\ast s}=g_\ast=106.75$ for universe early enough, and in our case, right after the electroweak phase transition it is slightly smaller~\cite{Husdal:2016haj}. In Fig.~\ref{fig:dfdxs}, we show the numerical result and compare it to the thermal differential partition functions. The overall normalization of the freeze-in differential partition function is determined by reproducing the relic abundance; here we choose a normalization similar in scale to the thermal ones for convenience of comparison.

The most useful feature for us is that on the large $y_\phi=\frac{p_\phi}{T}$ side the partition function is also suppressed exponentially, closely following the thermal distributions. All freeze-in partition function calculations found similar behavior due to the inevitable Boltzmann suppression of either parent particle or the center of mass energy, for the freeze-in process of both UV type and IR type~\cite{Roland:2016gli}, and regardless of whether the portal achieves thermal equilibrium~\cite{Konig:2016dzg,Biswas:2016iyh}. In small $y_\phi$ regime, we can see a deviation from thermal distributions, and different models can give quite different shapes~\cite{Konig:2016dzg,Biswas:2016iyh}. However the bulk of FIDM particles resides on the $\frac{p_\phi}{T}\gtrsim1$ region and dictates the MPS shape, so in general, the low momentum side is not important and can be approximately replaced by a thermal one such as the Fermi-Dirac one in the WDM model. We will see how well it works soon. A partition function that cannot be approximated by a single thermal distribution should be possible, in case that the multiple components with high and low momenta are comparable. However, such case is somewhat tuned in parameters, and the multiple components should be able to get approximated by multiple components of WDM. So in the following we will use this differential partition functions with $m_h=125$~GeV and $m_\phi=7$~keV for illustration, and the small shape difference for different $m_\phi$ value is ignored.

Further, we perform numerical analysis by solving the perturbation Boltzmann equation set for the evolution of the FS DM using \texttt{camb}~\cite{Lewis:1999bs}. The general formulation follows~\cite{Ma:1995ey} with the choice of synchronous gauge. The DM follows the standard massive neutrino perturbative Boltzmann equation
\begin{align}
\label{eq:massive neutrino}
\partial_\tau\Psi_0=&-\frac{\tilde{p}}{\tilde{E}}k\Psi_1+\frac{1}{6}\partial_\tau h\frac{d\ln f}{d\ln\tilde{p}},\\
\partial_\tau\Psi_1=&\frac{\tilde{p}}{\tilde{E}}k\Big(\Psi_0-\frac{2}{3}\Psi_2\Big),\\
\partial_\tau\Psi_2=&\frac{\tilde{p}}{\tilde{E}}k\Big(\frac{2}{5}\Psi_1\hspace{-0.2em}-\hspace{-0.2em}\frac{3}{5}\Psi_3\Big)\hspace{-0.2em}-\hspace{-0.2em}\Big(\frac{1}{15}\partial_\tau h\hspace{-0.2em}+\hspace{-0.2em}\frac{2}{5}\partial_\tau\eta\Big)\frac{d\ln f}{d\ln\tilde{p}},\\
\partial_\tau\Psi_\ell=&\frac{\tilde{p}}{\tilde{E}}k\frac{1}{2\ell+1}\Big(\ell\Psi_{\ell-1}-(\ell+1)\Psi_{\ell+1}\Big)\quad\ell\geq3,\\
\partial_\tau\Psi_\ell=&\frac{\tilde{p}}{\tilde{E}}k\Psi_{\ell-1}+\frac{\ell+1}{\tau}\Psi_\ell\qquad\text{as truncation}.
\end{align}
Here $\Psi_\ell$ is the fractional perturbation from the unperturbed partition function $f(\tilde{p})$ and the subscript labels the harmonic mode, $\tau$ is conformal time and $\tilde{p}=ap$ and $\tilde{E}=\sqrt{\tilde{p}^2+a^2m^2}$ are comoving momentum and energy respectively, $k$ is the interested Fourier mode, and $h$ and $\eta$ are the metric perturbation in synchronous gauge. Different from the redshifted Fermi-Dirac distribution of benchmark WDM model, the afore calculated differential partition function is inputted in the $\frac{d\ln f}{d\ln\tilde{p}}$ terms in the $\ell=0$ and $\ell=2$ equations.

\begin{figure}[t!]
\includegraphics[scale=0.7]{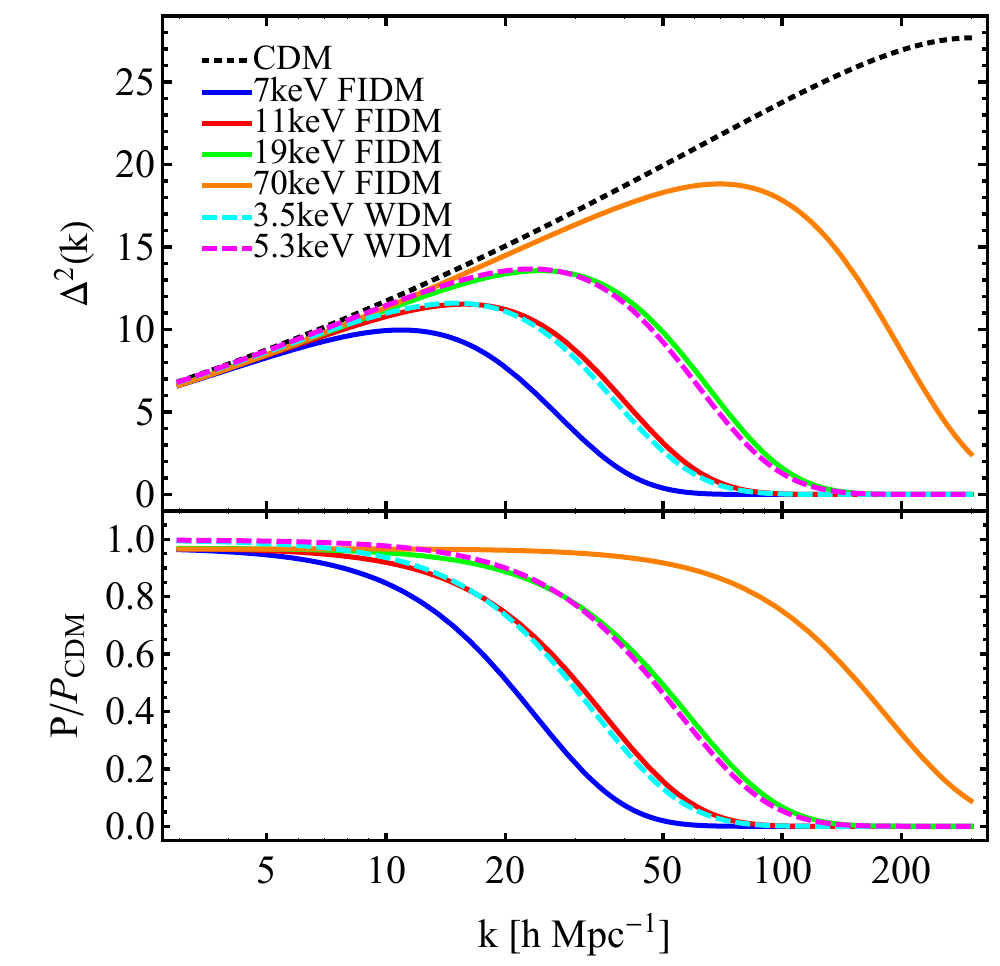}
\caption{MPS of our light scalar FIDM model (upper panel) and the ratios to the CDM MPS (bottom panel), for $g_{\ast s}=106.75$ (for simplicity) and several FIDM masses. Also shown are the WDM ones at the $2\sigma$ Lyman-$\alpha$ bound of $m_\text{WDM}=3.5$~keV ($5.3$~keV)~\cite{Irsic:2017ixq}. The FIDM model $m_\phi=11$~keV ($19$~keV) is chosen that the characteristic FS velocity ($\frac{\tilde{p}}{\tilde{E}}$ in Eq.~\ref{eq:massive neutrino}) matches that of the $3.5$~keV ($5.3$~keV) thermal WDM model. The remaining tiny difference is caused by the nonthermal partition function $\frac{d\ln f}{d\ln\tilde{p}}$.}
\label{fig:MPSFI}
\end{figure}
In Fig.~\ref{fig:MPSFI} we show the MPS ($\Delta(k)=\frac{k^3}{2\pi^2}P(k)$) of FIDM, in comparison with benchmark WDM ones. While tuning the characteristic FS velocities to be the same, we can see that the MPS of light DM is indeed not sensitive to the detailed shape of the unperturbed differential partition function such as the ones shown in Fig.~\ref{fig:dfdxs} (at least in the case that the new partition function also contains a Boltzmann suppression on the large comoving momentum side)\footnote{In the context of small scale MPS suppression due to FS, another known class of unperturbed partition function arises in the late decaying DM model~\cite{Huo:2011nz}. However, the MPS can also be very similar to the thermal WDM one in shape while tuning the FS velocity. Note that an arbitrary unperturbed partition function may not match to a realistic particle physics model. On the other hand, if the small scale MPS suppression is not due to FS but due to acoustic oscillation in the dark sector, then the MPS shape can be quite different~\cite{Cyr-Racine:2015ihg,Huo:2017vef}.}, and rather indistinguishable from that of a thermal distribution. The more suppressed MPS of the $7$~keV FIDM model means that it is excluded. This exclusion can be understood by simply considering the FS velocity, which is the factor $\frac{\tilde{p}}{\tilde{E}}$ in Eq.~\ref{eq:massive neutrino}. Viewing the Boltzmann equation set, when one ignores the small difference caused by non-thermal partition function such as that in our FIDM example, the FS velocity is the only factor which controls the MPS, and the more frequently used FS length~\cite{Kolb:1990vq} is just its integration under certain fixed Hubble behavior from the bigbang till matter-radiation equality (or later)
\begin{equation}
\label{eq:FSL}
\lambda=\int_0^{t_\text{eq}}v(t)\frac{dt}{a(t)}=\int_0^{a_\text{eq}}v(a)\frac{da}{a^2H(a)}.
\end{equation}
The convenient FS velocity to compare is the FS velocity extrapolated till today assuming no late virialization in structure formation,
\begin{align*}
\label{eq:FSV}
\frac{T}{\langle p\rangle}\langle v_0\rangle=&\frac{T_\text{WDM}}{m_\text{WDM}}=\Big(\frac{\rho_c\Omega_\nu}{m_\text{WDM}^4}\frac{2\pi^2}{3\zeta(3)}\Big)^{\frac{1}{3}}
,\\
=&\frac{T_\text{FIDM}}{m_\text{FIDM}}=\frac{T_0}{m_\text{FIDM}}\Big(\frac{g_{\ast s0}}{g_{\ast s\text{FIDM}}}\Big)^{\frac{1}{3}}
,
\end{align*}
where for massless fermion $\frac{T}{\langle p\rangle}=\frac{180\zeta(3)}{7\pi^4}\approx0.32$ and here for simplicity for all light DM we use this value and ignore the difference caused by partition function, and $g_{\ast s\text{FIDM}}$ is the effective $g_{\ast s}$ when the FIDM is produced. For the reason seen in the next section, we advocate it rather than the FS length. Then the matching above suggests that a WDM mass bound can be generally recast into an FIDM mass bound
\begin{equation}
\label{eq:weakscattering}
m_\text{FIDM}\geq24.3\Big(\frac{106.75}{g_{\ast s\text{FIDM}}}\Big)^{\frac{1}{3}}\Big(\frac{m_\text{WDM}}{6.3~\text{keV}}\Big)^{\frac{4}{3}}~\text{keV}.
\end{equation}
Here we have set the reference WDM bound to be the most recent one from the stellar stream observation~\cite{Banik:2019smi}.

\stepcounter{sec}
{\bf \Roman{sec}. The Strong Self-Scattering Limit.\;}
The random motion of a particle due to frequent (self) scattering is known as Brownian motion~\cite{Brown:1827}. The early DM Brownian motion in the universe will reduce the total FS length, by disabling FS in a specific direction from the bigbang to the decoupling of self-scattering. Here we show that such a non-free-streaming Brownian early stage will not work very effectively in protecting the small scale MPS from suppression.

As a simple approximation, we will set the FS velocity $\frac{\tilde{p}}{\tilde{E}}$ in Eq.~\ref{eq:massive neutrino} to be zero before the Brownian stage decoupling, which results in the same behavior with the Cold DM (CDM). Moreover, since elastic scattering will not statistically change the DM velocity with the latter only subject to redshift, after decoupling the FS velocity will retake the corresponding value as if there is no self-scattering. This is different from the old self-interacting DM MPS calculation~\cite{deLaix:1995vi}, since here we do not have heating in the $3\to2$ process. Such working approximation is easily implemented in \texttt{camb}~\cite{Lewis:1999bs}, by turning off the DM velocity before a specific decoupling time.

\begin{figure}[t!]
\includegraphics[scale=0.7]{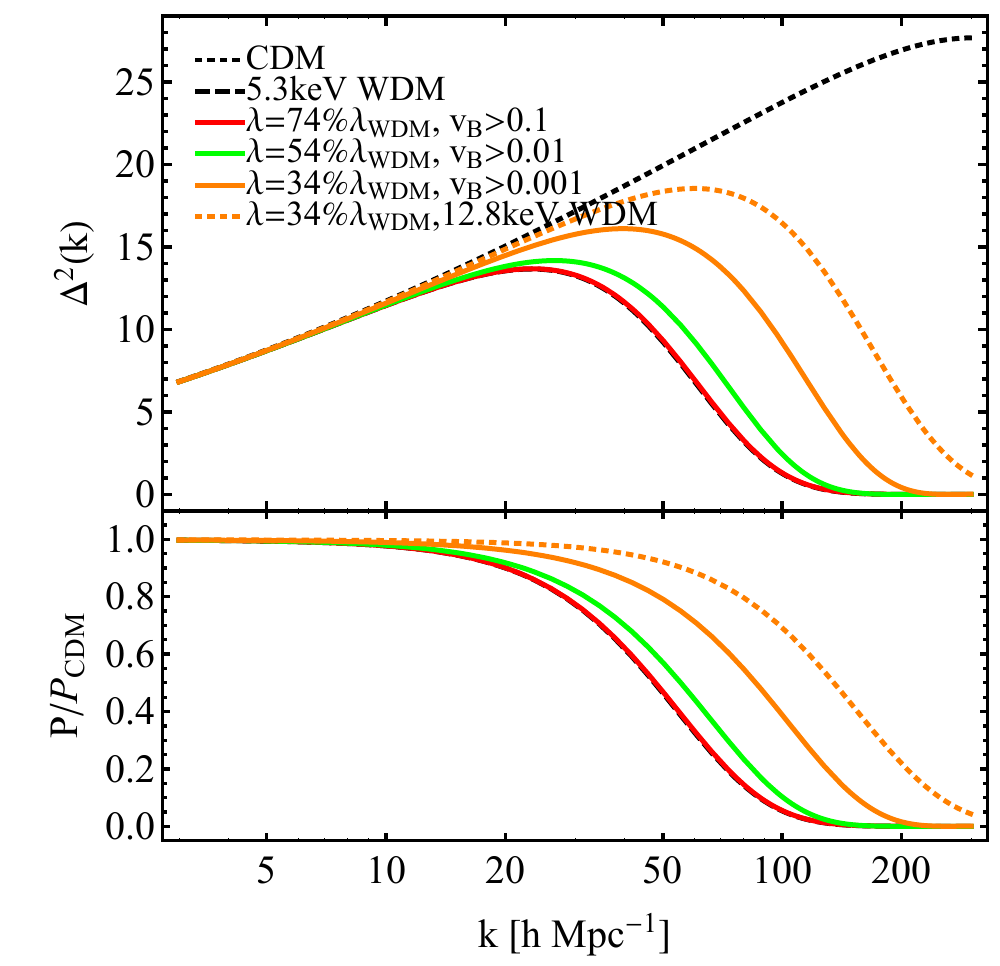}
\caption{MPS based on $m_\text{WDM}=5.3~\text{keV}$ WDM with an early Brownian stage during which DM macroscopic FS is forbidden (upper panel), and the ratios to the CDM MPS (bottom panel). Here we define such a stage by the DM velocity $v_\text{B}$ in it. The legend $\lambda$s originally defined in Eq.~\ref{eq:FSL} are now calculated with the decouplings of this stage as the starting points and compared to the benchmark thermal $5.3$~keV WDM one $\lambda_\text{WDM}$. Also shown is a MPS with the same total $\lambda$ as the last one in the series, but without such a stage and its reduction is due to a larger mass.}
\label{fig:vBcutFS}
\end{figure}
In Fig.~\ref{fig:vBcutFS} in the above approximation we calculate the MPS for the models with an early Brownian stage, which shares the same late-stage FS velocity with the $5.3$~keV thermal WDM model~\cite{Irsic:2017ixq}. Similar to the previous analysis, although the equilibrium partition may not be a Fermi-Dirac one but be a bosonic one or distorted by chemical potential, we can still use the WDM one for convenience, and the difference should be negligible. We can see that the MPS with early Brownian stages are almost indistinguishable from the benchmark thermal WDM model without an early Brownian stage, unless the decoupling happens very late till when DM becomes very non-relativistic. If the FS length is calculated from Eq.~\ref{eq:FSL} with the bigbang as the lower integration bound replaced by the Brownian decoupling, then actually till as high as a $50\%$ (the green curve) reduction is achieved, one can only see a mild protection against FS suppression. Moreover, the reduction of FS length in this way will not correspond to a unique MPS with the same total FS length but no Brownian early stage, for example matching the FS length corresponding to a Brownian decoupling DM velocity $v=10^{-3}~c$ we get $m_\text{WDM}=12.8$~keV for the thermal WDM model, but the MPS suppression for the latter is seen at a significantly smaller scale.

Apparently, the total FS length is not an excellent way to parameterize this effect. It is because the perturbation growth is not linear with the accumulation of FS length but the late time dominates. Even if an early Brownian stage prevents FS and makes the perturbation grow as effectively as CDM, such growth is driven by the dominant radiation-gravitation couple, and the magnitude is small at early times. And after the Brownian decoupling, the perturbation growth which is fast and dominant for CDM will still be erased by FS for light DM. On the other hand, if the self-scattering decoupling is pushed to the late side all the way till the matter radiation equilibrium or even later, we can indeed see the reduction of the FS effect.

\begin{figure}[t!]
\includegraphics[scale=0.6]{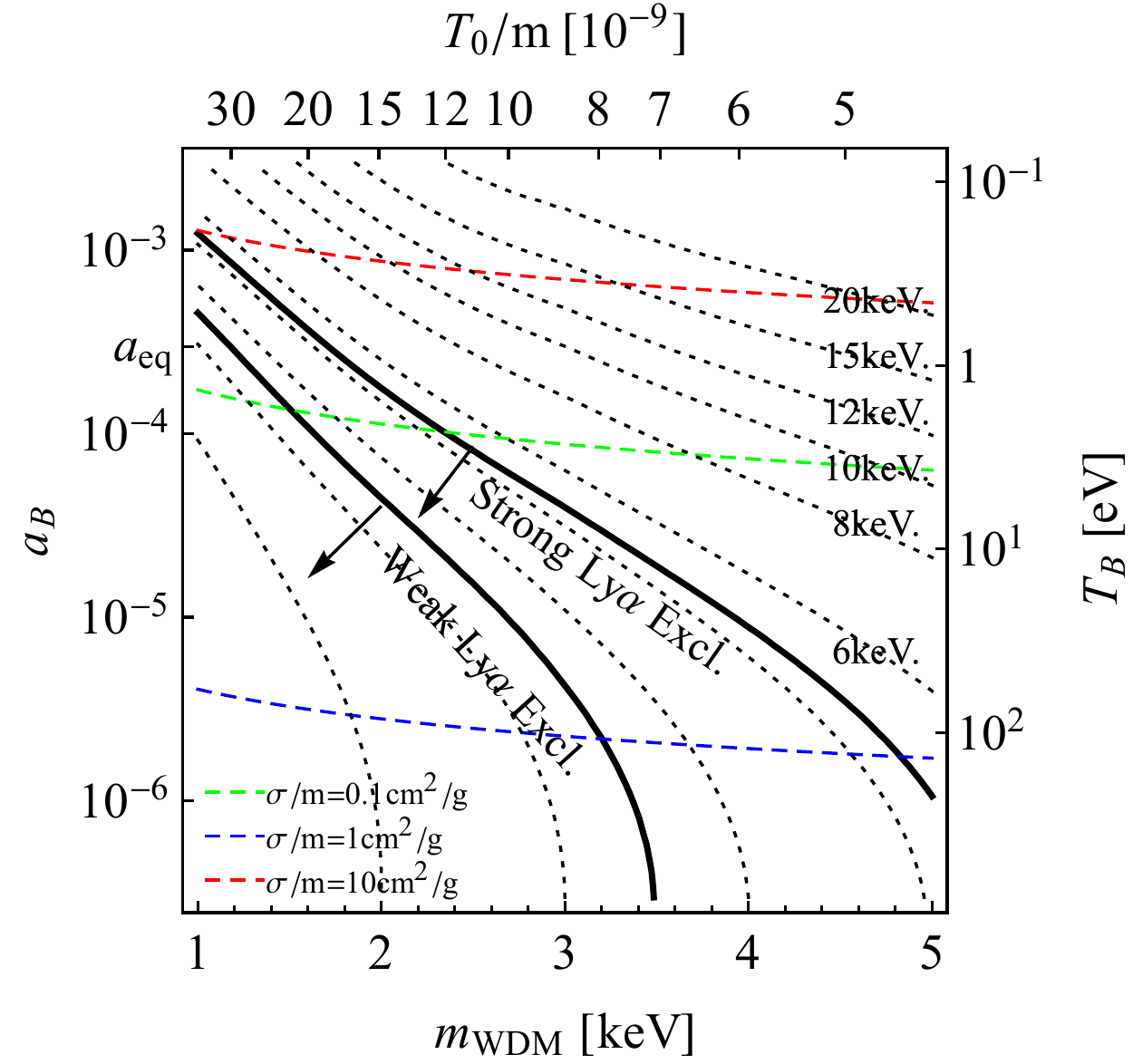}
\caption{Recast $2\sigma$ Lyman-$\alpha$ bounds~\cite{Irsic:2017ixq} based on the $\delta A$ criteria on the extrapolated FS $T_0/m$ vs.~Brownian decoupling $a_\text{B}$ plane, which are equivalent to thermal WDM mass and Brownian decoupling temperature $T_\text{B}$ respectively. A few more (dotted) contours are also shown. For early Brownian decoupling, they go back to the benchmark thermal WDM values. The more recent stellar stream $+$ dwarf satellite count WDM bound is $m_\text{WDM}=6.3~\text{keV}$~\cite{Banik:2019smi}, but now the $\delta A$ criteria is not applicable and the assumed self-scattering here may introduce new smearing effect to small scales, so the real exclusion contour will be different from the $m_\text{WDM}=6.3~\text{keV}$ one. We have also plotted three $a_\text{B}$ curves (colored and dashed) as functions of extrapolated FS velocity $T_0/m$ for reference, for different DM self-scattering cross section.}
\label{fig:mWDMvsaB}
\end{figure}
To better quantify the effect of the Brownian stage for MPS, we treat the Brownian motion decoupling scale factor $a_\text{B}$ as a convenient parameter, and scan the two parameters and compare with the current Lyman-$\alpha$ bound of thermal WDM. Given the DM self-scattering cross section which is usually expressed as $\frac{\sigma}{m}$, $a_\text{B}$ can be solved through the $\Gamma\sim H$ decoupling criteria, namely the equation
\begin{equation}
\frac{\rho_c\Omega_\chi}{a_\text{B}^3}\frac{\sigma}{m}\frac{\langle p\rangle}{T}\frac{T_0}{a_\text{B}m}=H_0\sqrt{\frac{\Omega_{\chi+b}}{a_\text{B}^3}+\frac{\Omega_{\gamma+\nu}}{a_\text{B}^4}+\Omega_\Lambda}
\end{equation}
For a DM self scattering cross section $\frac{\sigma}{m}\sim\text{cm}^2/\text{g}$~\cite{Tulin:2017ara} and an extrapolated FS velocity $\frac{T_0}{m}\sim10^{-8}$, interestingly $a_\text{B}$ is around the matter radiation equilibrium and the reduction of FS is visible. The result is shown in Fig.~\ref{fig:mWDMvsaB}. Constraints are made based on the $\delta A$ criteria~\cite{Murgia:2017lwo}: we first calculate the one-dimensional power spectrum of $P_\text{1d}(k)=\frac{1}{2\pi}\int_k^\infty k'dk'P(k')$ from three-dimensional MPS $P(k)$, then calculate $\delta A=1-\int_{k_\text{min}}^{k_\text{max}}dk \big(P_\text{1d}(k)/P_\text{1d}^{\text{CDM}}(k)\big)/(k_\text{max}-k_\text{min})$ with $k_\text{min}=0.5h/\text{Mpc}$ and $k_\text{max}=20h/\text{Mpc}$ and compare it to the $3.5$~keV ($5.3$~keV) thermal WDM model result. Such criteria should not be directly applied to the more recent WDM bound of $6.3$~keV based on stellar stream $+$ dwarf satellite count~\cite{Banik:2019smi}, and similar recast work is beyond the current work.

\stepcounter{sec}
{\bf \Roman{sec}. Summary.\;}
In this letter we have explored several aspects of the light freeze-in dark matter model, of its partition function determined by the freeze-in process without sufficient self-scattering, of the numerically calculated matter power spectrum shape and its similarity to a warm dark matter one at the same free-streaming velocity, and oppositely that whether a fast self-scattering or Brownian early stage will result in a protection from free-streaming suppression, and how such protection can be parameterized. We can see a generic similarity to the warm dark matter free-streaming effect, characterized by the extrapolated free streaming velocity, as well as the need of introducing Brownian decoupling time as the second parameter in the latter case. The light freeze-in dark matter model is also severely constrained by recast WDM bound, currently with a bound of about $24$~keV.

Acknowledgments: We are grateful to useful discussions with Haipeng An and Hai-Bo Yu.


\begin{thebibliography}{37}
\expandafter\ifx\csname natexlab\endcsname\relax\def\natexlab#1{#1}\fi
\expandafter\ifx\csname bibnamefont\endcsname\relax
  \def\bibnamefont#1{#1}\fi
\expandafter\ifx\csname bibfnamefont\endcsname\relax
  \def\bibfnamefont#1{#1}\fi
\expandafter\ifx\csname citenamefont\endcsname\relax
  \def\citenamefont#1{#1}\fi
\expandafter\ifx\csname url\endcsname\relax
  \def\url#1{\texttt{#1}}\fi
\expandafter\ifx\csname urlprefix\endcsname\relax\def\urlprefix{URL }\fi
\providecommand{\bibinfo}[2]{#2}
\providecommand{\eprint}[2][]{\url{#2}}

\bibitem[{\citenamefont{Aghanim et~al.}(2018)}]{Aghanim:2018eyx}
\bibinfo{author}{\bibfnamefont{N.}~\bibnamefont{Aghanim}} \bibnamefont{et~al.}
  (\bibinfo{collaboration}{Planck}) (\bibinfo{year}{2018}),
  \eprint{1807.06209}.

\bibitem[{\citenamefont{Kolb and Turner}(1990)}]{Kolb:1990vq}
\bibinfo{author}{\bibfnamefont{E.~W.} \bibnamefont{Kolb}} \bibnamefont{and}
  \bibinfo{author}{\bibfnamefont{M.~S.} \bibnamefont{Turner}},
  \bibinfo{journal}{Front. Phys.} \textbf{\bibinfo{volume}{69}},
  \bibinfo{pages}{1} (\bibinfo{year}{1990}).

\bibitem[{\citenamefont{Aprile et~al.}(2018)}]{Aprile:2018dbl}
\bibinfo{author}{\bibfnamefont{E.}~\bibnamefont{Aprile}} \bibnamefont{et~al.}
  (\bibinfo{collaboration}{XENON}), \bibinfo{journal}{Phys. Rev. Lett.}
  \textbf{\bibinfo{volume}{121}}, \bibinfo{pages}{111302}
  (\bibinfo{year}{2018}), \eprint{1805.12562}.

\bibitem[{\citenamefont{Cui et~al.}(2017)}]{Cui:2017nnn}
\bibinfo{author}{\bibfnamefont{X.}~\bibnamefont{Cui}} \bibnamefont{et~al.}
  (\bibinfo{collaboration}{PandaX-II}), \bibinfo{journal}{Phys. Rev. Lett.}
  \textbf{\bibinfo{volume}{119}}, \bibinfo{pages}{181302}
  (\bibinfo{year}{2017}), \eprint{1708.06917}.

\bibitem[{\citenamefont{Ackermann et~al.}(2017)}]{TheFermi-LAT:2017vmf}
\bibinfo{author}{\bibfnamefont{M.}~\bibnamefont{Ackermann}}
  \bibnamefont{et~al.} (\bibinfo{collaboration}{Fermi-LAT}),
  \bibinfo{journal}{Astrophys. J.} \textbf{\bibinfo{volume}{840}},
  \bibinfo{pages}{43} (\bibinfo{year}{2017}), \eprint{1704.03910}.

\bibitem[{\citenamefont{Boveia and Doglioni}(2018)}]{Boveia:2018yeb}
\bibinfo{author}{\bibfnamefont{A.}~\bibnamefont{Boveia}} \bibnamefont{and}
  \bibinfo{author}{\bibfnamefont{C.}~\bibnamefont{Doglioni}},
  \bibinfo{journal}{Ann. Rev. Nucl. Part. Sci.} \textbf{\bibinfo{volume}{68}},
  \bibinfo{pages}{429} (\bibinfo{year}{2018}), \eprint{1810.12238}.

\bibitem[{\citenamefont{Ellis et~al.}(1984)\citenamefont{Ellis, Kim, and
  Nanopoulos}}]{Ellis:1984eq}
\bibinfo{author}{\bibfnamefont{J.~R.} \bibnamefont{Ellis}},
  \bibinfo{author}{\bibfnamefont{J.~E.} \bibnamefont{Kim}}, \bibnamefont{and}
  \bibinfo{author}{\bibfnamefont{D.~V.} \bibnamefont{Nanopoulos}},
  \bibinfo{journal}{Phys. Lett.} \textbf{\bibinfo{volume}{145B}},
  \bibinfo{pages}{181} (\bibinfo{year}{1984}).

\bibitem[{\citenamefont{McDonald}(2002)}]{McDonald:2001vt}
\bibinfo{author}{\bibfnamefont{J.}~\bibnamefont{McDonald}},
  \bibinfo{journal}{Phys. Rev. Lett.} \textbf{\bibinfo{volume}{88}},
  \bibinfo{pages}{091304} (\bibinfo{year}{2002}), \eprint{hep-ph/0106249}.

\bibitem[{\citenamefont{Choi and Roszkowski}(2005)}]{Choi:2005vq}
\bibinfo{author}{\bibfnamefont{K.-Y.} \bibnamefont{Choi}} \bibnamefont{and}
  \bibinfo{author}{\bibfnamefont{L.}~\bibnamefont{Roszkowski}},
  \bibinfo{journal}{AIP Conf. Proc.} \textbf{\bibinfo{volume}{805}},
  \bibinfo{pages}{30} (\bibinfo{year}{2005}), \eprint{hep-ph/0511003}.

\bibitem[{\citenamefont{McDonald and Sahu}(2009)}]{McDonald:2008ua}
\bibinfo{author}{\bibfnamefont{J.}~\bibnamefont{McDonald}} \bibnamefont{and}
  \bibinfo{author}{\bibfnamefont{N.}~\bibnamefont{Sahu}},
  \bibinfo{journal}{Phys. Rev.} \textbf{\bibinfo{volume}{D79}},
  \bibinfo{pages}{103523} (\bibinfo{year}{2009}), \eprint{0809.0247}.

\bibitem[{\citenamefont{Hall et~al.}(2010)\citenamefont{Hall, Jedamzik,
  March-Russell, and West}}]{Hall:2009bx}
\bibinfo{author}{\bibfnamefont{L.~J.} \bibnamefont{Hall}},
  \bibinfo{author}{\bibfnamefont{K.}~\bibnamefont{Jedamzik}},
  \bibinfo{author}{\bibfnamefont{J.}~\bibnamefont{March-Russell}},
  \bibnamefont{and} \bibinfo{author}{\bibfnamefont{S.~M.} \bibnamefont{West}},
  \bibinfo{journal}{JHEP} \textbf{\bibinfo{volume}{03}}, \bibinfo{pages}{080}
  (\bibinfo{year}{2010}), \eprint{0911.1120}.

\bibitem[{\citenamefont{Bernal et~al.}(2017)\citenamefont{Bernal, Heikinheimo,
  Tenkanen, Tuominen, and Vaskonen}}]{Bernal:2017kxu}
\bibinfo{author}{\bibfnamefont{N.}~\bibnamefont{Bernal}},
  \bibinfo{author}{\bibfnamefont{M.}~\bibnamefont{Heikinheimo}},
  \bibinfo{author}{\bibfnamefont{T.}~\bibnamefont{Tenkanen}},
  \bibinfo{author}{\bibfnamefont{K.}~\bibnamefont{Tuominen}}, \bibnamefont{and}
  \bibinfo{author}{\bibfnamefont{V.}~\bibnamefont{Vaskonen}},
  \bibinfo{journal}{Int. J. Mod. Phys.} \textbf{\bibinfo{volume}{A32}},
  \bibinfo{pages}{1730023} (\bibinfo{year}{2017}), \eprint{1706.07442}.

\bibitem[{\citenamefont{Knapen et~al.}(2017)\citenamefont{Knapen, Lin, and
  Zurek}}]{Knapen:2017xzo}
\bibinfo{author}{\bibfnamefont{S.}~\bibnamefont{Knapen}},
  \bibinfo{author}{\bibfnamefont{T.}~\bibnamefont{Lin}}, \bibnamefont{and}
  \bibinfo{author}{\bibfnamefont{K.~M.} \bibnamefont{Zurek}},
  \bibinfo{journal}{Phys. Rev.} \textbf{\bibinfo{volume}{D96}},
  \bibinfo{pages}{115021} (\bibinfo{year}{2017}), \eprint{1709.07882}.

\bibitem[{\citenamefont{Ibe et~al.}(2018)\citenamefont{Ibe, Nakano, Shoji, and
  Suzuki}}]{Ibe:2017yqa}
\bibinfo{author}{\bibfnamefont{M.}~\bibnamefont{Ibe}},
  \bibinfo{author}{\bibfnamefont{W.}~\bibnamefont{Nakano}},
  \bibinfo{author}{\bibfnamefont{Y.}~\bibnamefont{Shoji}}, \bibnamefont{and}
  \bibinfo{author}{\bibfnamefont{K.}~\bibnamefont{Suzuki}},
  \bibinfo{journal}{JHEP} \textbf{\bibinfo{volume}{03}}, \bibinfo{pages}{194}
  (\bibinfo{year}{2018}), \eprint{1707.07258}.

\bibitem[{\citenamefont{Babu and Mohapatra}(2014)}]{Babu:2014pxa}
\bibinfo{author}{\bibfnamefont{K.~S.} \bibnamefont{Babu}} \bibnamefont{and}
  \bibinfo{author}{\bibfnamefont{R.~N.} \bibnamefont{Mohapatra}},
  \bibinfo{journal}{Phys. Rev.} \textbf{\bibinfo{volume}{D89}},
  \bibinfo{pages}{115011} (\bibinfo{year}{2014}), \eprint{1404.2220}.

\bibitem[{\citenamefont{Biswas et~al.}(2018)\citenamefont{Biswas, Choubey,
  Covi, and Khan}}]{Biswas:2017ait}
\bibinfo{author}{\bibfnamefont{A.}~\bibnamefont{Biswas}},
  \bibinfo{author}{\bibfnamefont{S.}~\bibnamefont{Choubey}},
  \bibinfo{author}{\bibfnamefont{L.}~\bibnamefont{Covi}}, \bibnamefont{and}
  \bibinfo{author}{\bibfnamefont{S.}~\bibnamefont{Khan}},
  \bibinfo{journal}{JCAP} \textbf{\bibinfo{volume}{1802}}, \bibinfo{pages}{002}
  (\bibinfo{year}{2018}), \eprint{1711.00553}.

\bibitem[{\citenamefont{Heeba et~al.}(2018)\citenamefont{Heeba, Kahlhoefer, and
  Stocker}}]{Heeba:2018wtf}
\bibinfo{author}{\bibfnamefont{S.}~\bibnamefont{Heeba}},
  \bibinfo{author}{\bibfnamefont{F.}~\bibnamefont{Kahlhoefer}},
  \bibnamefont{and} \bibinfo{author}{\bibfnamefont{P.}~\bibnamefont{Stocker}}
  (\bibinfo{year}{2018}), \eprint{1809.04849}.

\bibitem[{\citenamefont{de~Vega et~al.}(2014)\citenamefont{de~Vega, Salucci,
  and Sanchez}}]{deVega:2013jfy}
\bibinfo{author}{\bibfnamefont{H.~J.} \bibnamefont{de~Vega}},
  \bibinfo{author}{\bibfnamefont{P.}~\bibnamefont{Salucci}}, \bibnamefont{and}
  \bibinfo{author}{\bibfnamefont{N.~G.} \bibnamefont{Sanchez}},
  \bibinfo{journal}{Mon. Not. Roy. Astron. Soc.}
  \textbf{\bibinfo{volume}{442}}, \bibinfo{pages}{2717} (\bibinfo{year}{2014}),
  \eprint{1309.2290}.

\bibitem[{\citenamefont{An et~al.}(2018)\citenamefont{An, Huo, and
  Liu}}]{An:2018nvz}
\bibinfo{author}{\bibfnamefont{H.}~\bibnamefont{An}},
  \bibinfo{author}{\bibfnamefont{R.}~\bibnamefont{Huo}}, \bibnamefont{and}
  \bibinfo{author}{\bibfnamefont{W.}~\bibnamefont{Liu}} (\bibinfo{year}{2018}),
  \eprint{1812.05699}.

\bibitem[{\citenamefont{Roland and Shakya}(2017)}]{Roland:2016gli}
\bibinfo{author}{\bibfnamefont{S.~B.} \bibnamefont{Roland}} \bibnamefont{and}
  \bibinfo{author}{\bibfnamefont{B.}~\bibnamefont{Shakya}},
  \bibinfo{journal}{JCAP} \textbf{\bibinfo{volume}{1705}}, \bibinfo{pages}{027}
  (\bibinfo{year}{2017}), \eprint{1609.06739}.

\bibitem[{\citenamefont{Konig et~al.}(2016)\citenamefont{Konig, Merle, and
  Totzauer}}]{Konig:2016dzg}
\bibinfo{author}{\bibfnamefont{J.}~\bibnamefont{Konig}},
  \bibinfo{author}{\bibfnamefont{A.}~\bibnamefont{Merle}}, \bibnamefont{and}
  \bibinfo{author}{\bibfnamefont{M.}~\bibnamefont{Totzauer}},
  \bibinfo{journal}{JCAP} \textbf{\bibinfo{volume}{1611}}, \bibinfo{pages}{038}
  (\bibinfo{year}{2016}), \eprint{1609.01289}.

\bibitem[{\citenamefont{Biswas and Gupta}(2017)}]{Biswas:2016iyh}
\bibinfo{author}{\bibfnamefont{A.}~\bibnamefont{Biswas}} \bibnamefont{and}
  \bibinfo{author}{\bibfnamefont{A.}~\bibnamefont{Gupta}},
  \bibinfo{journal}{JCAP} \textbf{\bibinfo{volume}{1703}}, \bibinfo{pages}{033}
  (\bibinfo{year}{2017}), \bibinfo{note}{[Addendum: JCAP1705,no.05,A02(2017)]},
  \eprint{1612.02793}.

\bibitem[{\citenamefont{Bae et~al.}(2018)\citenamefont{Bae, Kamada, Liew, and
  Yanagi}}]{Bae:2017dpt}
\bibinfo{author}{\bibfnamefont{K.~J.} \bibnamefont{Bae}},
  \bibinfo{author}{\bibfnamefont{A.}~\bibnamefont{Kamada}},
  \bibinfo{author}{\bibfnamefont{S.~P.} \bibnamefont{Liew}}, \bibnamefont{and}
  \bibinfo{author}{\bibfnamefont{K.}~\bibnamefont{Yanagi}},
  \bibinfo{journal}{JCAP} \textbf{\bibinfo{volume}{1801}}, \bibinfo{pages}{054}
  (\bibinfo{year}{2018}), \eprint{1707.06418}.

\bibitem[{\citenamefont{Spergel and Steinhardt}(2000)}]{Spergel:1999mh}
\bibinfo{author}{\bibfnamefont{D.~N.} \bibnamefont{Spergel}} \bibnamefont{and}
  \bibinfo{author}{\bibfnamefont{P.~J.} \bibnamefont{Steinhardt}},
  \bibinfo{journal}{Phys. Rev. Lett.} \textbf{\bibinfo{volume}{84}},
  \bibinfo{pages}{3760} (\bibinfo{year}{2000}), \eprint{astro-ph/9909386}.

\bibitem[{\citenamefont{Tulin and Yu}(2018)}]{Tulin:2017ara}
\bibinfo{author}{\bibfnamefont{S.}~\bibnamefont{Tulin}} \bibnamefont{and}
  \bibinfo{author}{\bibfnamefont{H.-B.} \bibnamefont{Yu}},
  \bibinfo{journal}{Phys. Rept.} \textbf{\bibinfo{volume}{730}},
  \bibinfo{pages}{1} (\bibinfo{year}{2018}), \eprint{1705.02358}.

\bibitem[{\citenamefont{Yaguna}(2011)}]{Yaguna:2011qn}
\bibinfo{author}{\bibfnamefont{C.~E.} \bibnamefont{Yaguna}},
  \bibinfo{journal}{JHEP} \textbf{\bibinfo{volume}{08}}, \bibinfo{pages}{060}
  (\bibinfo{year}{2011}), \eprint{1105.1654}.

\bibitem[{\citenamefont{Husdal}(2016)}]{Husdal:2016haj}
\bibinfo{author}{\bibfnamefont{L.}~\bibnamefont{Husdal}},
  \bibinfo{journal}{Galaxies} \textbf{\bibinfo{volume}{4}}, \bibinfo{pages}{78}
  (\bibinfo{year}{2016}), \eprint{1609.04979}.

\bibitem[{\citenamefont{Lewis et~al.}(2000)\citenamefont{Lewis, Challinor, and
  Lasenby}}]{Lewis:1999bs}
\bibinfo{author}{\bibfnamefont{A.}~\bibnamefont{Lewis}},
  \bibinfo{author}{\bibfnamefont{A.}~\bibnamefont{Challinor}},
  \bibnamefont{and} \bibinfo{author}{\bibfnamefont{A.}~\bibnamefont{Lasenby}},
  \bibinfo{journal}{Astrophys. J.} \textbf{\bibinfo{volume}{538}},
  \bibinfo{pages}{473} (\bibinfo{year}{2000}), \eprint{astro-ph/9911177}.

\bibitem[{\citenamefont{Ma and Bertschinger}(1995)}]{Ma:1995ey}
\bibinfo{author}{\bibfnamefont{C.-P.} \bibnamefont{Ma}} \bibnamefont{and}
  \bibinfo{author}{\bibfnamefont{E.}~\bibnamefont{Bertschinger}},
  \bibinfo{journal}{Astrophys. J.} \textbf{\bibinfo{volume}{455}},
  \bibinfo{pages}{7} (\bibinfo{year}{1995}), \eprint{astro-ph/9506072}.

\bibitem[{\citenamefont{Irsic et~al.}(2017)}]{Irsic:2017ixq}
\bibinfo{author}{\bibfnamefont{V.}~\bibnamefont{Irsic}} \bibnamefont{et~al.},
  \bibinfo{journal}{Phys. Rev.} \textbf{\bibinfo{volume}{D96}},
  \bibinfo{pages}{023522} (\bibinfo{year}{2017}), \eprint{1702.01764}.

\bibitem[{\citenamefont{Banik et~al.}(2019)\citenamefont{Banik, Bovy, Bertone,
  Erkal, and de~Boer}}]{Banik:2019smi}
\bibinfo{author}{\bibfnamefont{N.}~\bibnamefont{Banik}},
  \bibinfo{author}{\bibfnamefont{J.}~\bibnamefont{Bovy}},
  \bibinfo{author}{\bibfnamefont{G.}~\bibnamefont{Bertone}},
  \bibinfo{author}{\bibfnamefont{D.}~\bibnamefont{Erkal}}, \bibnamefont{and}
  \bibinfo{author}{\bibfnamefont{T.~J.~L.} \bibnamefont{de~Boer}}
  (\bibinfo{year}{2019}), \eprint{1911.02663}.

\bibitem[{\citenamefont{Brown}(1828)}]{Brown:1827}
\bibinfo{author}{\bibfnamefont{R.}~\bibnamefont{Brown}},
  \bibinfo{journal}{privately circulated}  (\bibinfo{year}{1828}).

\bibitem[{\citenamefont{de~Laix et~al.}(1995)\citenamefont{de~Laix, Scherrer,
  and Schaefer}}]{deLaix:1995vi}
\bibinfo{author}{\bibfnamefont{A.~A.} \bibnamefont{de~Laix}},
  \bibinfo{author}{\bibfnamefont{R.~J.} \bibnamefont{Scherrer}},
  \bibnamefont{and} \bibinfo{author}{\bibfnamefont{R.~K.}
  \bibnamefont{Schaefer}}, \bibinfo{journal}{Astrophys. J.}
  \textbf{\bibinfo{volume}{452}}, \bibinfo{pages}{495} (\bibinfo{year}{1995}),
  \eprint{astro-ph/9502087}.

\bibitem[{\citenamefont{Murgia et~al.}(2017)\citenamefont{Murgia, Merle, Viel,
  Totzauer, and Schneider}}]{Murgia:2017lwo}
\bibinfo{author}{\bibfnamefont{R.}~\bibnamefont{Murgia}},
  \bibinfo{author}{\bibfnamefont{A.}~\bibnamefont{Merle}},
  \bibinfo{author}{\bibfnamefont{M.}~\bibnamefont{Viel}},
  \bibinfo{author}{\bibfnamefont{M.}~\bibnamefont{Totzauer}}, \bibnamefont{and}
  \bibinfo{author}{\bibfnamefont{A.}~\bibnamefont{Schneider}}
  (\bibinfo{year}{2017}), \eprint{1704.07838}.

\bibitem[{\citenamefont{Huo}(2011)}]{Huo:2011nz}
\bibinfo{author}{\bibfnamefont{R.}~\bibnamefont{Huo}}, \bibinfo{journal}{Phys.
  Lett.} \textbf{\bibinfo{volume}{B701}}, \bibinfo{pages}{530}
  (\bibinfo{year}{2011}), \eprint{1104.4094}.

\bibitem[{\citenamefont{Cyr-Racine et~al.}(2016)\citenamefont{Cyr-Racine,
  Sigurdson, Zavala, Bringmann, Vogelsberger, and
  Pfrommer}}]{Cyr-Racine:2015ihg}
\bibinfo{author}{\bibfnamefont{F.-Y.} \bibnamefont{Cyr-Racine}},
  \bibinfo{author}{\bibfnamefont{K.}~\bibnamefont{Sigurdson}},
  \bibinfo{author}{\bibfnamefont{J.}~\bibnamefont{Zavala}},
  \bibinfo{author}{\bibfnamefont{T.}~\bibnamefont{Bringmann}},
  \bibinfo{author}{\bibfnamefont{M.}~\bibnamefont{Vogelsberger}},
  \bibnamefont{and} \bibinfo{author}{\bibfnamefont{C.}~\bibnamefont{Pfrommer}},
  \bibinfo{journal}{Phys. Rev.} \textbf{\bibinfo{volume}{D93}},
  \bibinfo{pages}{123527} (\bibinfo{year}{2016}), \eprint{1512.05344}.

\bibitem[{\citenamefont{Huo et~al.}(2018)\citenamefont{Huo, Kaplinghat, Pan,
  and Yu}}]{Huo:2017vef}
\bibinfo{author}{\bibfnamefont{R.}~\bibnamefont{Huo}},
  \bibinfo{author}{\bibfnamefont{M.}~\bibnamefont{Kaplinghat}},
  \bibinfo{author}{\bibfnamefont{Z.}~\bibnamefont{Pan}}, \bibnamefont{and}
  \bibinfo{author}{\bibfnamefont{H.-B.} \bibnamefont{Yu}},
  \bibinfo{journal}{Phys. Lett.} \textbf{\bibinfo{volume}{B783}},
  \bibinfo{pages}{76} (\bibinfo{year}{2018}), \eprint{1709.09717}.

\end{thebibliography}
\end{document}